\documentstyle[amssymb,prd,aps,floats]{revtex}
\begin{document}
\draft

\input epsf \renewcommand{\topfraction}{0.8} 
\twocolumn[\hsize\textwidth\columnwidth\hsize\csname 
@twocolumnfalse\endcsname

\title{Some remarks on Oscillating Inflation}
\author{V. C\'{a}rdenas\thanks{%
E-mail: vcardena@lauca.usach.cl} and G. Palma\thanks{%
E-mail: gpalma@lauca.usach.cl}}
\address{Departamento de F\'{i}sica, Universidad de Santiago\\
Casilla 307, Correo 2, Santiago, Chile.}
\date{\today}
\maketitle

\begin{abstract}
In a recent paper Damour and Mukhanov describe a scenario where inflation
may continue during the oscillatory phase. This effect is possible because
the scalar field spends a significant fraction of each period of oscillation
on the upper part of the potential. Such additional period of inflation
could push perturbations after the slow roll regime to observable scales.
Although in this work we show that the small region of the Damour-Mukhanov
parameter q gives the main contribution to oscillating inflation, it was not
satisfactory understood until now. Furthermore, it gives an expression for
the energy density spectrum of perturbations, which is well behaved in the
whole physical range of q .
\end{abstract}


%\input epsf \renewcommand{\topfraction}{0.8} 
%@twocolumnfalse\endcsname

\vskip2pc] 
%%%%%%%%%%%%%%%%%%%%%%%%%%%%%%%%%%%%%%%%%%%%%%%%%%%%%%%%%%%%%%%%%%%%%%%%

\section{INTRODUCTION}

Nowadays inflation is a widely accepted element of the early cosmology \cite
{inf}. It gives the possibility of solving many of the shortcomings of the
standard hot big bang model and provides the source for the early energy
density fluctuations responsible of the large scale structure of the
universe observed today. Although there are many models of inflation, the
underlying physical ideas are well established. These are characterized by a
period of ``slow roll'' evolution of a scalar field (called inflaton) toward
the vacuum potential. During this period the field changes very slowly, so
that the kinetic energy $\dot{\varphi}^{2}/2$ remains smaller than its
potential energy $V(\varphi )$. The energy density associated to the scalar
field acts as a ``cosmological constant'' term, allowing a period of quasi
exponential expansion of the scale factor. When the period of inflation
ends, the scalar field $\varphi $ start a phase of rapid coherent
oscillations around the vacuum.

Very recently (\cite{dm,lm}) it has been pointed out that inflation can
persist during the coherent oscillations of the inflation field phase. This
exciting result is possible when the inflaton potential verifies a simple
constrain of curvature far from the core convex part, where the inflaton
field can roll slowly. The efficiency of this phenomena could have important
implications for \ GUT scale baryogenesis \cite{klr}. In fact, as suggested
by Damour and Mukhanov (\cite{dm}),it can be expected that due to the
increase of the oscillation frequency, there is the possibility to generate
massive particles heavier than $\sim 10^{16}GeV$.

In ref.\cite{dm} Damour and Mukhanov estimated the amount of inflation to be 
$\sim 10$ e-fold (powers of the scale factor). They argue that this effect
can be more efficient than the parametric resonance effect \cite{kls} for
the amplification of cosmological perturbations \cite{fi-bran}.\ In ref.\cite
{lm} Liddle and Mazumdar showed that Mukhanov et al. overestimated the
number of e-fold because they have used a slow-roll definition of this
object. In their paper, Liddle and Mazumdar found an analytical expression
for the number of e-fold of inflation using the appropriate definition
finding a number of $\sim 3$ e-fold concluding that this effect is not very
efficient. The study of adiabatic perturbations in this phase has been made
by Taruya \cite{taru}. He found a poor amplification in the case of a single
scalar field model but anticipated an enormous amplification for multi-field
systems.

In this letter we review the problem. In particular we find that the
analytical expressions used to compare with the numerical estimation are not
well defined in the $q\sim 0$ region and propose a way to correct these
analytical estimations. Furthermore, with this result we study the evolution
of the scalar field finding total agreement with the conclusions of ref.\cite
{lm} for $q>0.2$, but a remarkable different result for small $q$. For this
region, the initial conditions are very important. We find that $q\sim 0$
gives the leading contribution for oscillating inflation and the dominant
part in the amplification of the fluctuations.

The letter is organized as follow; first we describe briefly the
Damour-Mukhanov model. Then, we make some comments about the initial
conditions for this phenomenon and later we propose an improved expression,
valid for the leading region of $q$, which is our main contribution.

\section{BASIC EQUATIONS}

Now we shall restrict ourselves to models of inflation driven by a single
scalar field. The equations are 
\begin{equation}
\ddot{\varphi}+3H\dot{\varphi}+V,_{\varphi }=0,  \label{ec1}
\end{equation}

\begin{equation}
H^{2}=\kappa ^{2}(\frac{1}{2}\dot{\varphi}^{2}+V),  \label{ec2}
\end{equation}
Here $H=\dot{a}/a$ is the Hubble parameter, $a$ is the scale factor of the
universe and $\kappa ^{2}=8\pi /3M_{p}^{2}$ with $M_{p}=1.2\cdot 10^{19}Gev$
the Planck mass. During the oscillatory phase of $\varphi $ we have two time
scales; the inverse of the frequency $\omega ^{-1}$ of oscillations of $%
\varphi $ and the inverse of the rate of expansion $H^{-1}$. If the limit $%
\omega \gg H$ is taken we can neglect terms proportional to $H$ in the
equations. So from (1) we can integrate to obtain, 
\begin{equation}
\rho =\frac{1}{2}\dot{\varphi}^{2}+V=cte=V_{m},  \label{ta1}
\end{equation}
where $V_{m}=V(\varphi _{m})$ is the maximum value of $V(\varphi )$ in each
oscillation when the field reaches the maximum value $\varphi _{m}$. From
this relation we obtain the period of a single oscillation, $%
T=4\int_{0}^{\varphi _{m}}d\varphi \left[ 2(V_{m}-V(\varphi )\right] ^{\frac{%
1}{2}}.$When $\omega \ll H$ we can define an adiabatic average index $\gamma 
$ by $\gamma =\left\langle (\rho +p)/\rho \right\rangle ,$where the bracket
means $\left\langle ...\right\rangle =T^{-1}\int_{0}^{T}...dt$. Equations
(1,2) can be re-written in the fluid form 
\begin{equation}
\dot{\rho}=-3H(p+\rho ),  \label{ec3}
\end{equation}

\begin{equation}
\frac{\ddot{a}}{a}=-\frac{1}{3}(\rho +3p),  \label{ec4}
\end{equation}
then from the definition of $\gamma $ and eqns.(\ref{ec3},\ref{ec4}) we have
several ways to compute the adiabatic index 
\begin{equation}
\gamma =\frac{\left\langle \dot{\varphi}^{2}\right\rangle }{V_{m}}=\frac{%
\left\langle \varphi V,_{\varphi }\right\rangle }{V_{m}}=2(1-\frac{%
\left\langle V\right\rangle }{V_{m}}).  \label{g}
\end{equation}
Because $p=(\gamma -1)\rho $ and (\ref{ec4}) we have a superluminal
expansion $\ddot{a}>0$ when $\gamma <2/3$. From the last two relations in
eqn.(\ref{g}) the inequality $\gamma <2/3$ leads to 
\begin{equation}
\left\langle V-\varphi V,_{\varphi }\right\rangle >0.  \label{rm}
\end{equation}

\section{THE DAMOUR-MUKHANOV MODEL}

Until now everything has been done for an arbitrary potential, but from now
on we shall consider the potential 
\begin{equation}
V(\varphi )=\frac{A}{q}\left[ \left( \frac{\varphi ^{2}}{\varphi _{c}^{2}}%
+1\right) ^{q/2}-1\right] ,  \label{pot}
\end{equation}
where $q$ is a dimensionless parameter, $A=[$mass$]^{4}$ is a constant and $%
\varphi _{c}=[$mass$]$ determines the size of the convex core of $V(\varphi
) $. We assume for a while that $\varphi _{c}$ marks the end of {\it %
oscillating inflation}. The analysis made in ref.\cite{dm} works well far
from the core of the potential. Further, the limit $\varphi \gg \varphi _{c}$
of eqn.(\ref{pot}) was written as: 
\begin{equation}
V(\varphi )\simeq \frac{A}{q}\left( \frac{\varphi }{\varphi _{c}}\right)
^{q}.  \label{pot1}
\end{equation}
In this case, the adiabatic index can be computed exactly given \cite{mt} by 
\begin{equation}
\gamma =\frac{2q}{q+2},  \label{gama}
\end{equation}
so, from the inequality $\gamma <2/3$ we note that to hold inflation during
the oscillatory phase we must have $q<1$. By using eqn.(\ref{gama}) in eqn.(%
\ref{ec3}) we obtain $\dot{\rho}=-3H\gamma \rho $ and together with eqn.(\ref
{ec2}) we have 
\begin{equation}
a\propto t^{2/3\gamma }=t^{(q+2)/3q},  \label{ee1}
\end{equation}
\begin{equation}
\varphi _{m}\propto t^{-2/q}\propto a^{-6/(q+2)},  \label{ee2}
\end{equation}
\begin{equation}
\rho =V(\varphi _{m})\propto t^{-2}\propto a^{-6q/(q+2)},  \label{ee3}
\end{equation}
where $\varphi _{m}$ is the amplitude of the oscillations, $\varphi
_{c}<\varphi _{m}<\varphi _{s}$ and $\varphi _{s}$ is a typical value of $%
\varphi $ at the end of slow-roll inflation and the beginning of oscillating
inflation. To compute the number of e-fold of inflation during oscillating
inflation we can not use the standard expression $N=\ln (a_{f}/a_{i})$,
appropriate for the slow-roll stage, but the improved expression proposed in
ref.\cite{lm} 
\begin{equation}
\tilde{N}=\ln \frac{a_{f}H_{f}}{a_{i}H_{i}},  \label{ne}
\end{equation}
because in each oscillation, while the field spends time in the core region,
the universe continue their expansion so $H$ can vary. Then from (\ref{ec2})
and (\ref{ee3}) $H\propto a^{-3q/(q+2)}$ the product $aH\propto \varphi
_{m}^{(1-q)/3}$ and from (\ref{ne}) we obtain 
\begin{equation}
\tilde{N}\simeq \frac{1-q}{3}\left[ \ln \frac{qM_{p}}{\varphi _{c}}-2\right]
,  \label{ne2}
\end{equation}
where we have used $\varphi _{s}\sim qM_{p}/\sqrt{16}\pi $. In \cite{lm} the
numerical curves for $\varphi _{c}=10^{-6}M_{p}$ show that $\tilde{N}%
\lesssim 3$. Using the analytical expression (\ref{ne2}) we do not find
agreement for small values of $q$. However there is not a compelling reason
to believe in (\ref{ne2}) for small $q$.

\section{THE SMALL q-REGION}

To study the small $q$ region, we must use the correct limit $q\rightarrow 0$
of (\ref{pot}) which leads to 
\begin{equation}
V(\varphi )\simeq \frac{A}{2}\ln \left[ \left( \frac{\varphi }{\varphi _{c}}%
\right) ^{2}+1\right] ,  \label{pot2}
\end{equation}
so, if now we take the limit $\varphi \gg \varphi _{c}$ we obtain the
logarithmic potential $V(\varphi )\simeq A\ln (\varphi /\varphi _{c})$. A
very important fact to note from (\ref{pot1}) is that the limit $%
q\rightarrow 0$ does not exist. Of course, the expression (\ref{pot1}) is
wrong around the $q\sim 0$ region and the expressions derived from this are
ill-defined. But, some work has been done in this regard \cite{dm}. For the
logarithmic potential the adiabatic index is $\gamma =1/\ln (\varphi
_{m}/\varphi _{c})$, so from (\ref{ec3}) and (\ref{ec2}) we obtain 
\begin{equation}
a(t)\propto \exp \left[ -\frac{A}{2}(t_{end}-t)^{2}\right] ,  \label{ta3}
\end{equation}
but this form does not permit us to write an explicit expression for $\tilde{%
N}$ (see ref. \cite{lm}). Let us make some comments about this result.
Because $\gamma =1/\ln (\varphi _{m}/\varphi _{c})$ from (\ref{ec3}) we
obtain $\dot{\rho}=-3HA$, then $a\sim (\varphi _{m}/\varphi _{c})^{-1/3}$.
Moreover, from (\ref{ec2}) we have $H\propto \rho ^{1/2}\propto (\ln
(\varphi _{m}/\varphi _{c}))^{1/2}$, then to compute $\tilde{N}$ we should
evaluate the factor $(\ln (\varphi _{m}/\varphi _{c}))^{1/2}(\varphi
_{m}/\varphi _{c})^{-1/3}$ at the extremes $\varphi _{s}$ and $\varphi _{c}$%
, but this is not possible. The problem arises when $\varphi _{m}$ is chosen
close to $\varphi _{c}$ in a expression valid for $\varphi \gg \varphi _{c}$%
. Because (\ref{pot1}) is valid for $\varphi \gg \varphi _{c}$ too, the same
problem should appear in the calculation of $\tilde{N}$. In fact this is the
case but, to see that, we must include the constant term $A/q$ in (\ref{pot1}%
) or equivalently, write the limit correctly.

When we use the potential 
\begin{equation}
V(\varphi )\simeq \frac{A}{q}\left[ \left( \frac{\varphi }{\varphi _{c}}%
\right) ^{q}-1\right] ,  \label{pot3}
\end{equation}
the adiabatic index become time-dependent, satisfying the equation 
\begin{equation}
\gamma \rho =\frac{2q}{q+2}(\rho +\frac{A}{q}),  \label{ta4}
\end{equation}
where $\rho (t)=V(\varphi _{m}(t))=A[(\varphi _{m}(t)/\varphi
_{c})^{q}-1]/q. $ Replacing this in (\ref{ec3}) we obtain the same behavior
as in eqn.(\ref{ee2}). But when we calculate $H$ we obtain $H\propto \rho
^{1/2}\propto q^{-1/2}[(\varphi _{m}/\varphi _{c})^{q}-1]^{1/2}$, which is
not well defined at the point $\varphi _{m}=\varphi _{c}$. The same happens
to the e-fold number, as mentioned before.

The authors of \cite{lm} found $\varphi _{s}\sim qM_{p}/\sqrt{16}\pi $ using
the potential (\ref{pot1}). From this result they found a decrease of $%
\tilde{N}$ at decreasing values of $q$. This seems very strange because, for
smaller values of $q$ we obtain flatter potentials at $\varphi \gg \varphi
_{c}$, then $\varphi $ rolls slowly most of the time increasing the amount
of inflation. Furthermore, we are not safe of how they set the initial
conditions for $\varphi $ in their numerical analysis. In general, the
initial and final field configuration depends on the potential.

For $q$ close to cero, $\varphi _{s}$ is not proportional to $q$. In fact,
we know that $\varphi _{s}$ came from the saturation of the slow-roll
inequality $\left| V^{\prime }/V\right| <\sqrt{48\pi }/M_{p}$ using the
expression (\ref{pot3}) for the potential 
\begin{equation}
\left( \frac{\varphi _{s}}{\varphi _{c}}\right) ^{q-1}=\frac{\sqrt{48\pi }%
\varphi _{c}}{qM_{p}}\left[ \left( \frac{\varphi _{s}}{\varphi _{c}}\right)
^{q}-1\right] .  \label{recur}
\end{equation}
for the $\varphi \gg \varphi _{c}$ region. In Figure 1 we see the behavior
of $\varphi _{s}$ in terms of $q$ given by eqn.(\ref{recur}). Of course, in
the large $q$-region both curves agree. In order to illustrate this point
and compare with ref.\cite{lm} we use its expression given by eqn.(\ref{ne})
but instead of using $\varphi _{s}\sim qM_{p}/\sqrt{16}\pi $ we use the
numerical values of $\varphi _{s}$ obtained from eqn.(\ref{recur}). The
results are plotted in Figure 2. In the small $q$-region the field $\varphi
_{s}$ grows preventing the fall of $\tilde{N}$ predicted in ref.\cite{lm}.

Moreover, as we have anticipated before, the value of the field at the end
of oscillating inflation will have a $q$-dependence too. We know from ref. 
\cite{dm} that the intercept $U(\varphi )=V(\varphi )-\varphi V_{,\varphi }$%
, must be positive to hold oscillating inflation. Let us define $\varphi
_{f} $ to be the value of the inflaton field $\varphi $ at which $U(\varphi
_{f})=0$. This condition represents the end of inflation due to oscillation.
We need thus to compare $\varphi _{f}$ for different values of $q$.

If we take the potential (\ref{pot3}) and define $x=\varphi /\varphi _{c}$,
we obtain 
\begin{equation}
U(x)=\frac{A}{q}\left[ x^{q}\left( 1-q\right) -1\right] .  \label{ta6}
\end{equation}
From this equation we can extract a explicit expression for $\varphi _{f}$.
If we impose $U(x_{f})=0$ we obtain the value of the scalar inflaton field
at the end of this phase 
\begin{equation}
\varphi _{f}=\varphi _{c}\left( 1-q\right) ^{-1/q}.  \label{fif}
\end{equation}
Using the improved expression eqn.(\ref{ne}) for the e-fold number (ref.\cite
{lm}), but inserting the corrected values for $\varphi _{s}$ and $\varphi
_{f}$ given by eqns. (\ref{recur}) and (\ref{fif}) we obtain 
\begin{equation}
\tilde{N}\simeq \ln \left\{ \left( \frac{\varphi _{s}}{\varphi _{f}}\right)
^{(2+q)/6}\left[ \frac{(\varphi _{f}/\varphi _{c})^{q}-1}{(\varphi
_{s}/\varphi _{c})^{q}-1}\right] ^{1/2}\right\} .  \label{ta7}
\end{equation}
The corrected value for $\varphi _{f}$ leads to a even smaller amount of
inflation when comparing with the value obtained in ref.\cite{lm}. In figure
3, we plot eqn.(\ref{ta7}) and show the behavior of both effects combined.
Because $\varphi _{f}$ is greater than $\varphi _{c}$, the amount of
inflation is smaller than one obtained by Liddle et al.\cite{lm} in the
whole range of $q\in (0,1)$. Moreover, the correct values of $\varphi _{s}$
produce a positive contribution to $\tilde{N}$ in the small $q$-range, which
avoids the fall of $\tilde{N}$, as $q$ goes to smaller values, predicted by
Liddle et. al.\cite{lm}. Again, it is not possible to show the whole range
of $q$ because $q>\varphi _{c}/\varphi $ (see comments below eqn.(\ref{recur}%
)).

Because oscillating inflation adds e-folds of inflation after the slow-roll
regime, where the observed perturbations are generated, is possible that
this additional period of inflation could push perturbations to observable
scales. In order to obtain the required amplitude of density perturbations
and without imposing unphysical constraint on the potential, we should
compute the density perturbation spectrum for the model being studied.

In reference \cite{lm} an expression for this object was derived 
\[
\delta _{H}^{2}=\frac{512\pi }{75}\frac{A}{q^{3}M_{p}^{6}}\frac{\varphi
^{q+2}}{\varphi _{c}^{q}}. 
\]
However, this expression is not well defined close enough to zero. Using the
primordial density perturbation spectrum $\delta _{H}$ as was define in \cite
{lm} we obtain for eqn.(\ref{pot3}) 
\begin{equation}
\delta _{H}^{2}=\frac{512\pi }{75}\frac{A}{q^{3}M_{p}^{6}}\left\{ \frac{%
\left[ \left( \varphi /\varphi _{c}\right) ^{q}-1\right] ^{3}}{\left(
\varphi /\varphi _{c}\right) ^{2q}}\right\} ,  \label{pertur}
\end{equation}
which is well defined even for the small values of $q$ : 
\begin{equation}
\delta _{H}^{2}\approx \frac{512\pi }{75}\frac{A}{M_{p}^{6}}\ln ^{3}\left( 
\frac{\varphi }{\varphi _{c}}\right) .  \label{pertur2}
\end{equation}
Because the COBE satellite require $\delta _{H}\approx 2\cdot 10^{-5}$, in
the $q=0$ case the amplitude of the potential for $\varphi _{c}=10^{-6}M_{p}$
gives $A^{1/4}\sim 2\cdot 10^{-3}M_{p}$, which is a typical number for
inflationary models.

\section{SUMMARY}

As a summary, we have made some corrections about how to compute the e-fold
number, which accounts for the amount of inflation during the oscillatory
phase. In particular we note that previous studies are not accurate because
they are not valid close to the core of the potential $\varphi \sim \varphi
_{c}$. Thus, we make the analysis for the small $q$ region of the potential,
which has not been considered until now. Finally we find that, in order to
extract the correct amount of inflation during this phase, a very careful
definition of initial and final field configuration is needed. Our results
show that near $q\sim 0$ the e-fold number is maximal but it is still not
enough to be more efficient than the parametric resonant effect discussed in 
\cite{kls}.

\section*{Acknowledgments}

The authors want to thanks Sergio del Campo for helpful discussions. V.
C\'{a}rdenas wants to thanks CONICYT for support through a scholarship. G.
Palma was supported in part through Proyect FONDECYT 1980608 and Proyect
DICYT 049631PA. 
%%%%%%%%%%%%%%%%%%%%%%%%%%%%%%%%%%%%%%%%%%%%%%%%%%%%%%%%%%%%%%%%%%%%%%%% 

\newpage

\bigskip

\section{Figures caption}

{\bf Figure 1}:\qquad We plot the q dependence of $\varphi _{s}$. We see
that for $q>0.2$ both curves agree but for smaller values of $q$ the field $%
\varphi _{s}$ grows, preventing the fall of $\tilde{N}$.

{\bf Figure 2}:\qquad The e-fold number $\tilde{N}$ is shown as a function
of $q$, taking into account the behavior of $\varphi _{s}$ described for the
eqn. (20).

{\bf Figure 3}:\qquad The e-fold number $\tilde{N}$ is plotted vs $q$,
taking into account the combined effects: the behavior of $\varphi _{s}$
described by eqn. (20) and the definition of $\varphi _{f}$ discussed in the
text.

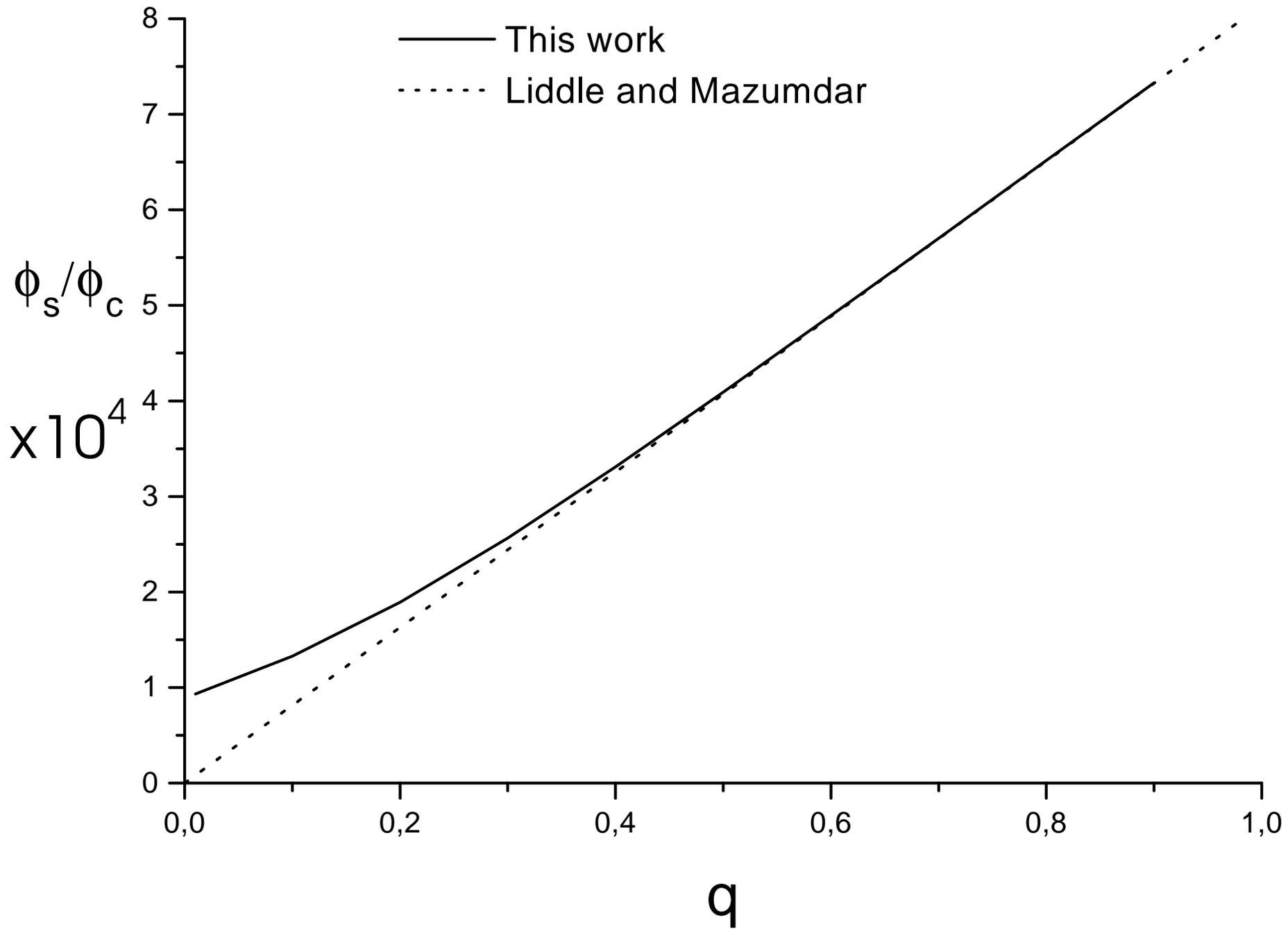
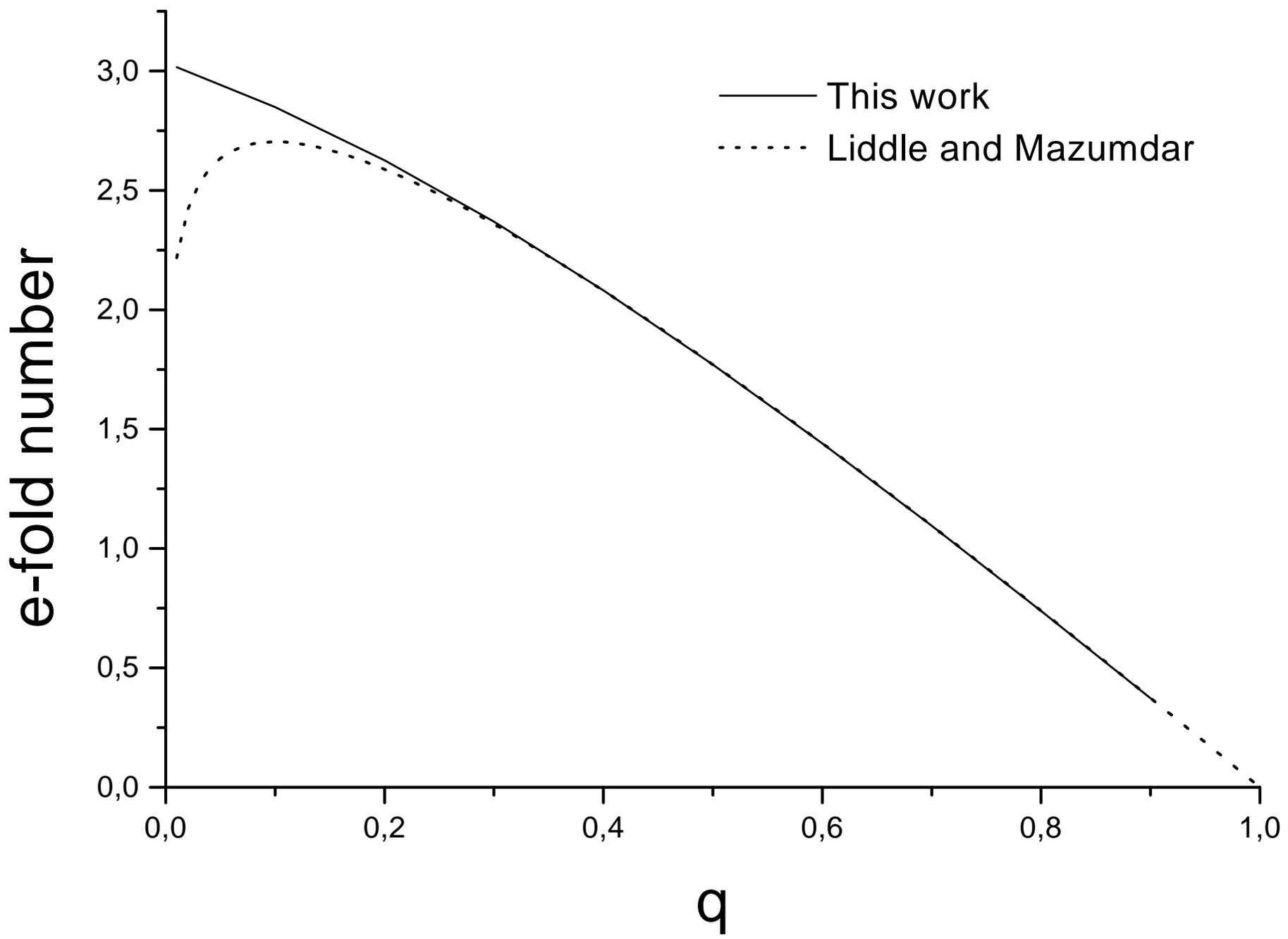
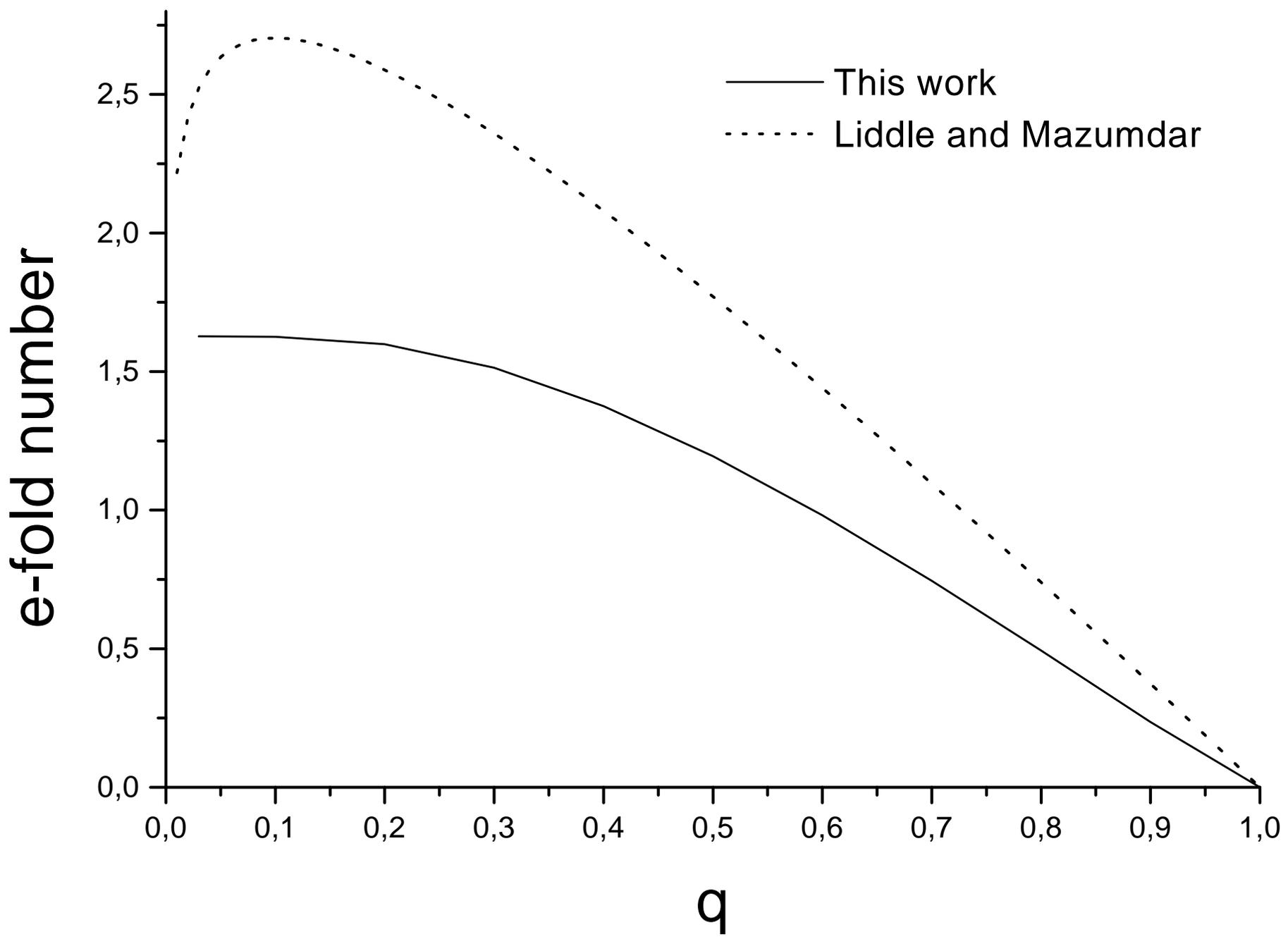
\end{document}